\documentclass[%
 reprint,
 showkeys,
 floatfix,
 amsmath,amssymb,
 aps,
]{revtex4-2}
\usepackage{graphicx}
\usepackage{dcolumn}
\usepackage{bm}
\usepackage{amsmath} 
\usepackage[utf8]{inputenc} 
\usepackage[export]{adjustbox}
\usepackage{comment}
\usepackage{booktabs}
\usepackage[colorlinks=true,linkcolor=blue,citecolor=blue,urlcolor=blue]{hyperref}
\usepackage{textgreek}
\usepackage{subfig}

\begin{document}

\title{First-principles study of hydrogen diffusion in polycrystalline Nickel }

\author{Bhanuj Jain\textsuperscript{a}}
\author{Alaa Olleak\textsuperscript{b}}
\author{Junyan He\textsuperscript{b}}
\author{Adarsh Chaurasia\textsuperscript{b}}
\author{Davide Di Stefano\textsuperscript{a,}}
\thanks{Corresponding author. \\ E-mail: \href{mailto:davide.distefano@synopsys.com}{davide.distefano@synopsys.com} (D. Stefano)}

\affiliation{\textsuperscript{a}Ansys UK Ltd, Cambridge, CB1 7EG, UK}
\affiliation{\textsuperscript{b}Ansys Inc, Canonsburg, PA 15317, USA}

\date{\today}

\begin{abstract} 
Hydrogen embrittlement in metals is strongly governed by hydrogen diffusion and trapping, yet predicting these effects in polycrystalline systems remains challenging. This work introduces a multiscale modeling framework that links atomistic energetics to continuum-scale transport. Migration barriers for bulk and grain-boundary environments, obtained from first-principles calculations, are used in kinetic Monte Carlo simulations to compute anisotropic effective diffusivities. These diffusivities are then incorporated into finite element models of polycrystalline microstructures, explicitly accounting for grain-boundary character and connectivity. The approach captures both fast-path and trapping effects without relying on empirical parameters and reproduces experimental trends for nickel, including the dependence of effective diffusivity on grain size and boundary type. This methodology provides a physically grounded route for predicting hydrogen transport in engineering alloys and can be extended to other materials and defect types.
\end{abstract}


\maketitle

\section{\label{sec:level1}Introduction}

Hydrogen transport in metals underlies a range of critical degradation mechanisms, including hydrogen embrittlement, delayed fracture, and stress corrosion cracking, which limit the performance and safety of structural materials in various industries~\cite{Takai2010_HDiffusion, Lynch2012_HEreview, YU_2024_review}. Understanding and predicting hydrogen mobility in complex engineering materials is therefore critical. In polycrystalline metals, overall hydrogen mobility is particularly complex as a result of a combination of bulk diffusion and interactions with microstructural features such as dislocations and grain boundaries (GB)~\cite{Herzig_2003, Kirchheim2007_HTrapping, LI202474}. These microstructural features can simultaneously trap hydrogen and act as fast diffusion pathways, resulting in strongly anisotropic and spatially heterogeneous transport.

First-principles and atomistic simulations have provided detailed insights into hydrogen behavior near specific defect configurations. Various authors~\cite{Du_2011_GB, HE20217589, MAI2021110283, HUANG2023107222, di2016first} have shown that hydrogen interaction with extended defects varies significantly depending on the material and atomic structure; some defects act as traps or barriers, while others enhance diffusion. However, such simulations are inherently limited in spatial extent and cannot directly capture hydrogen transport across micrometer-scale polycrystals.
To overcome this limitation, kinetic Monte Carlo (KMC) methods have been used to statistically homogenize atomistic transport behavior over larger scales~\cite{Du_2012, Zieb_2015, Kumar_2020}. 

On the other hand, continuum models for hydrogen diffusion typically rely on phenomenological trapping formulations, such as the McNabb–Foster and Oriani models~\cite{McNabbFoster1963, Oriani1970_trapping}, which describe hydrogen exchange between lattice sites and traps under equilibrium or kinetic assumptions. In these approaches, microstructural defects are represented using trap densities, binding energies, and exchange rates calibrated from experiments~\cite{Krom2000}. While effective at the engineering scale, such models typically assume that trapped hydrogen is immobile and do not resolve the local energetics or anisotropic transport behavior arising from the atomistic structure. These simplifications overlook the fact that extended defects, dislocations, interphases, and grain boundaries can act as fast diffusion pathways while trapping hydrogen, with transport rates comparable to or exceeding those in the bulk. This limitation is especially pronounced for grain boundaries, whose structural and chemical complexity often leads to strongly directional transport and concurrent trapping behavior~\cite{Zhou_2019, Li_2021antagonist}. Such effects cannot be captured by point-trap formulations but require microstructure-resolved continuum models that explicitly incorporate finite thickness and directional transport properties.

Over the past decades, many extensions of classical models have been proposed~\cite{CHEN2025789, Barrera2018_HydrogenReview, YU_2024_review}. These often rely on an increasing number of  parameters, e.g., trap densities, binding energies, and rate constants~\cite{LOPESPINTO2024105116, TAHA2001803, DADFARNIA201110141}. However, many of these parameters are not directly measurable~\cite{Marrani2025_ML_TDS}, and their effects on macroscopic diffusion behavior are often non-unique. As a result, the already challenging model calibration and validation~\cite{McNabbFoster1983_classic} becomes increasingly complex and potentially an ill-posed problem.

In this work, to bridge atomistic and continuum level modeling, we have used KMC to extract transport properties from atomistic data and transfer them into continuum models in a physically consistent manner. The resulting multiscale model provides quantitative predictions of effective hydrogen diffusivity as a function of grain size and grain-boundary character, offering a physically grounded pathway for linking defect structure to macroscopic transport behavior.

The remainder of this paper is organized as follows. Section~\ref{sec:methodology} describes the multiscale approach, including atomistic input, kinetic Monte Carlo simulations, and continuum finite element implementation. Section~\ref{sec:results} presents simulation results for representative grain-boundary configurations in nickel, highlighting the influence of microstructural features on effective hydrogen diffusivity. Section~\ref{sec:discussion} discusses the implications of the findings and the generalizability of the approach to other defect types. Finally, Section~\ref{sec:conclusions} summarizes the main conclusions and outlines directions for future work.

\section{\label{sec:methodology}Methodology}

The multiscale methodology developed in this work links atomistic diffusion energetics to continuum-scale transport through an intermediate KMC description. Migration barriers for bulk and grain-boundary environments are taken from literature density-functional theory (DFT) studies and are used as input for KMC simulations on a supercell representing a system of infinite parallel boundaries at a certain distance in a slab-model, similar to typical atomistic supercells. These simulations proved effective diffusivities within finite regions influenced by GB.

In the continuum finite element model, grain boundaries are represented as bands of prescribed thickness embedded within a representative volume element (RVE), while grain interiors are assigned bulk diffusivity. The key modeling assumption is that a finite region surrounding a grain boundary can be treated as an effective medium whose transport response is equivalent to that of a periodically repeated slab, provided that the continuum grain-boundary thickness matches the slab periodicity used in the KMC simulations. This correspondence allows the direct use of KMC-derived diffusivities as constitutive properties of the grain-boundary regions.

Details of the atomistic input, KMC simulations, and continuum model implementations are provided in the following subsections.

\subsection{\label{sec:GB_DFT}Atomistic data}
The atomistic input data used in this work were obtained from first-principles simulations of hydrogen in nickel.  
The diffusion parameters for the bulk lattice were taken from Di~Stefano \emph{et al.}~\cite{DiStefano2015_PRB92_224301}, while the GBs energetics and diffusivities were adopted from Di~Stefano \emph{et al.}~\cite{DiStefano2015_Acta98_306}.  
Both studies were based on DFT calculations, ensuring a consistent description of hydrogen migration in crystalline and interfacial regions.

In Ref.~\cite{DiStefano2015_PRB92_224301}, hydrogen diffusion in FCC Ni was investigated using the projector-augmented wave method within the GGA–PBE functional. Quantum effects, including zero-point vibrations and tunneling, were incorporated through a semiclassical transition-state formalism. 

Hydrogen segregation and diffusion at Ni grain boundaries were characterized in Ref.~\cite{DiStefano2015_Acta98_306}.  
That study examined representative coincident-site lattice boundaries, namely $\Sigma3\,(111)[110]$ (hereafter referred to as $\Sigma3$) and $\Sigma5\,(210)[001]$ ($\Sigma5$), using the same DFT framework as that used for the bulk.

Hydrogen binding and migration energies were computed for various interstitial sites within and near the boundary plane.

The study showed that the interaction of hydrogen with grain boundaries in nickel depends strongly on the boundary structure. The $\Sigma3$ GB exhibits a closed-packed interface structure and shows negligible hydrogen trapping; hydrogen is not trapped at the interface, and the boundary actually acts as a two-dimensional diffusion barrier rather than a fast pathway, with an energy barrier to cross the GB of about 0.55 eV.

In contrast, the more disordered $\Sigma5$ GB contains open interstitial sites that act as a two-dimensional trap with energy barriers to enter and escape the boundary of about 0.06 eV and 0.26 eV, respectively. Hydrogen diffusion along such disordered boundaries is significantly faster than in the bulk due to reduced effective migration barriers (approximately 0.27 eV).

In this work, the $\Sigma3$ and $\Sigma5$ are used as representatives of special and random GBs which are known to have markedly different effects on H diffusivity~\cite{OUDRISS20126814, BECHTLE20094148}.

The results collected were subsequently used as input for the kinetic Monte Carlo (KMC) model described in the following section.

\subsection{\label{sec:KMC}Kinetic Monte Carlo}

To bridge the atomistic energetics of hydrogen migration with microstructurally relevant diffusion behavior, a KMC approach was employed. KMC is a stochastic technique that reproduces the time evolution of systems governed by thermally activated events through probabilistic sampling of transition rates. Unlike molecular dynamics, which explicitly resolves atomic vibrations on the femtosecond scale, KMC advances the system through discrete diffusion events, thereby enabling access to macroscopic scales otherwise unreachable by direct atomistic simulation~\cite{BORTZ197510,GILLESPIE1976403,10.1063/1.1415500,10.1063/1.461138}.

\begin{figure}[!htbp]
    \centering
    \subfloat{
        \includegraphics[width=0.49\columnwidth, frame]{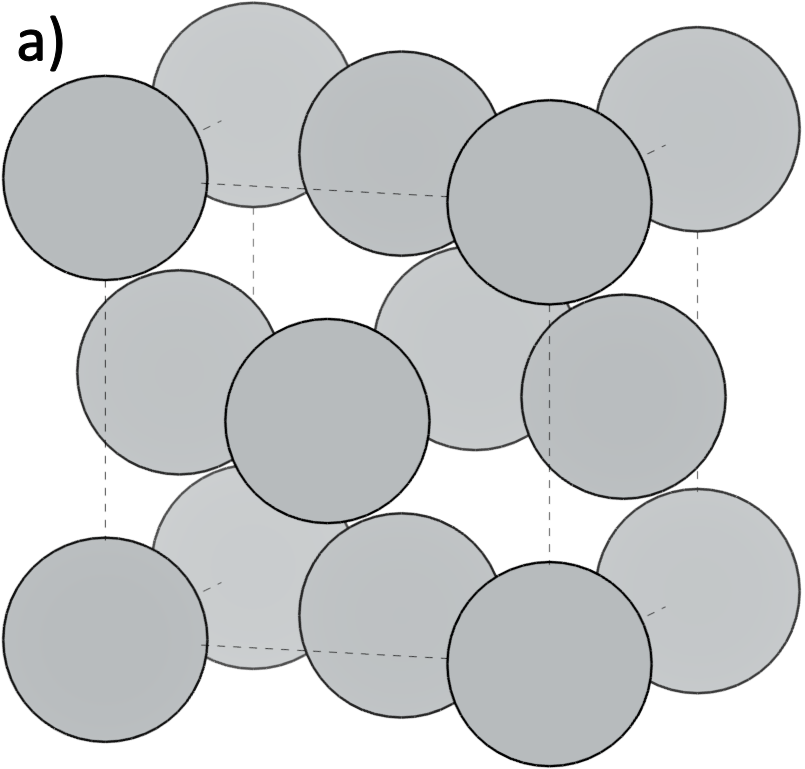}%
        \label{fig:atomistic_a}%
    }\hfill
    \subfloat{%
        \includegraphics[width=0.49\columnwidth, frame]{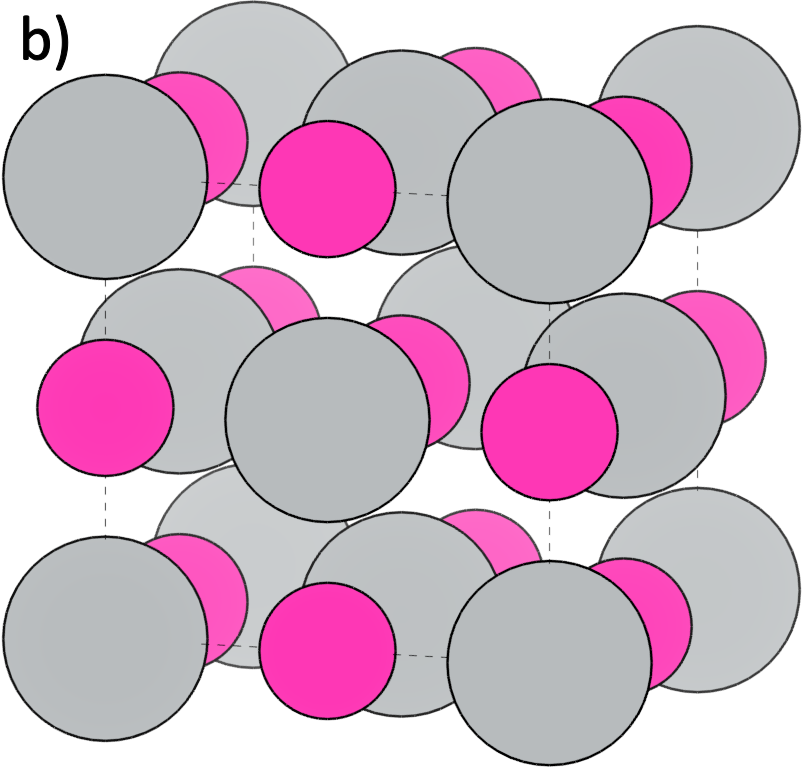}%
        \label{fig:atomistic_b}%
    }\\[0.5em] 
    
    \subfloat{%
        \includegraphics[height=0.47\columnwidth, frame]{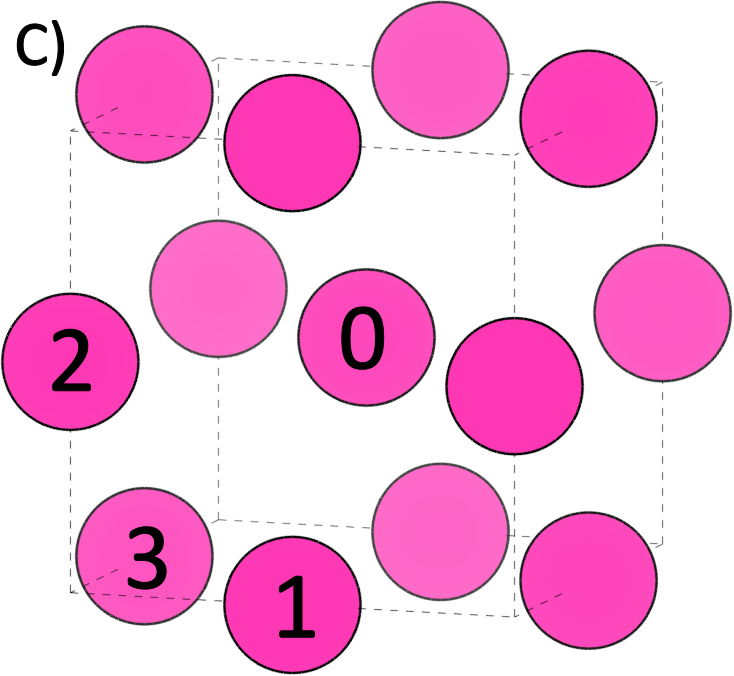}%
        \label{fig:atomistic_c}%
    }\hfill
    \subfloat{%
        \includegraphics[height=0.47\columnwidth, frame]{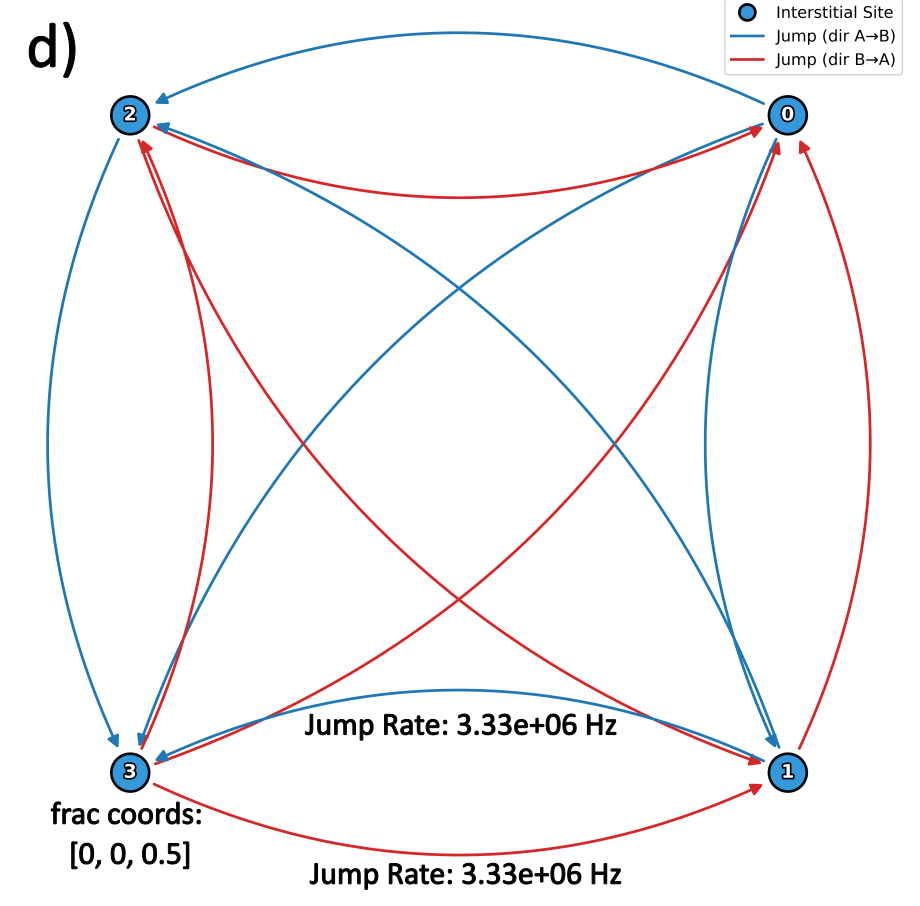}%
        \label{fig:atomistic_d}%
    }
    
    \caption{Workflow for converting a crystal structure into its KMC diffusion graph for bulk Ni. The starting point (a) is the Ni perfect lattice. The lattice is then (b) decorated with the relevant interstitial sites, octahedral in this case. The original Ni atoms are removed (c). Finally, (d) the diffusion lattice is constructed. Note, only in-cell jumps are represented, but periodic images are considered in the code}
    \label{fig:struct_to_graph}
\end{figure}

A custom graph-based KMC framework was implemented to model hydrogen diffusion across bulk and grain-boundary (GB) regions. The diffusion network was represented as a graph, with interstitial sites as nodes and possible migration paths as edges. Each edge connecting sites $i$ and $j$ was assigned an activation barrier $\Delta E_{ij}$ and an attempt frequency $\nu_{ij}$, obtained from first-principles energetics. The workflow to get from a crystal structure to such graphs is illustrated in Fig.~\ref{fig:struct_to_graph}. 

The graph representation allows for efficient incorporation of microstructural heterogeneity—such as interfaces and defect-rich regions—while maintaining a direct link to atomistic input.

\begin{figure}[!htbp]
	\centering
	\includegraphics[width=0.8\linewidth, frame]{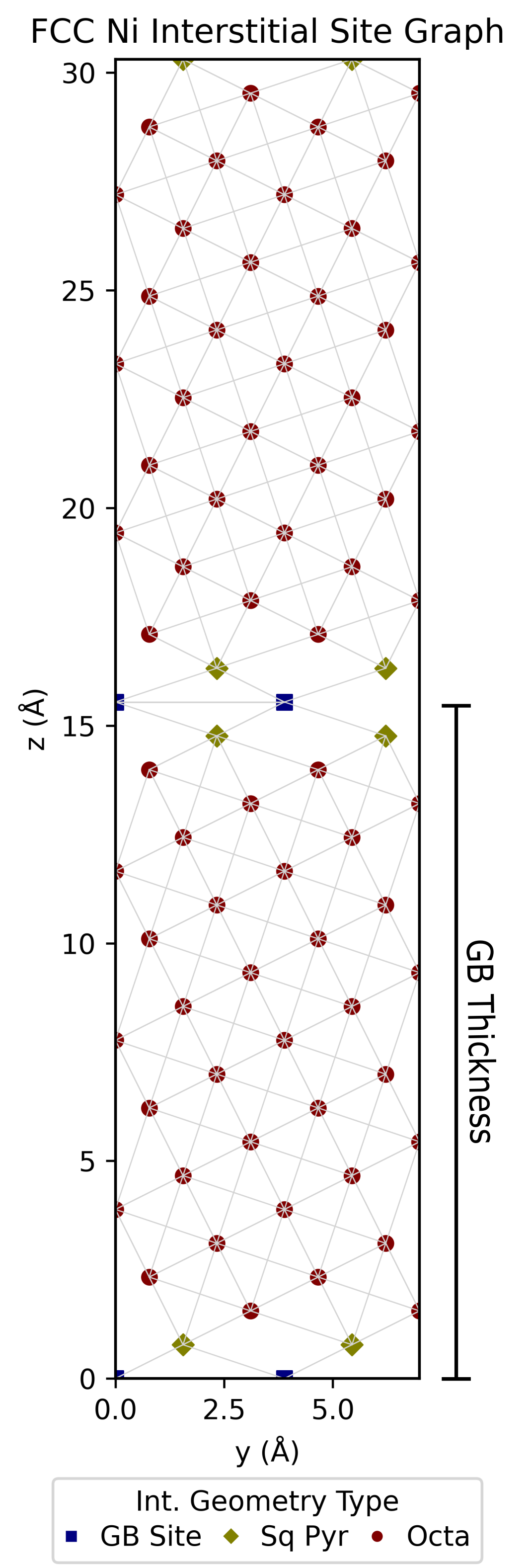}    
	\caption{Graph representation of interstitial diffusion sites in FCC Ni. Most of possible sites are Ni octahedral sites (red circles). Approaching the GB plane, octahedral site are distorted in to square-pyramidal sites (olive diamonds), and finally in the GB plane effective sites (blue squares) nodes. Grey lines represent the possible hydrogen migration paths.}
	\label{fig:GB_graph}
	\vspace{-2pt}
\end{figure}

\begin{figure}[!htbp]
	\centering
	\includegraphics[width=0.99\columnwidth, frame]{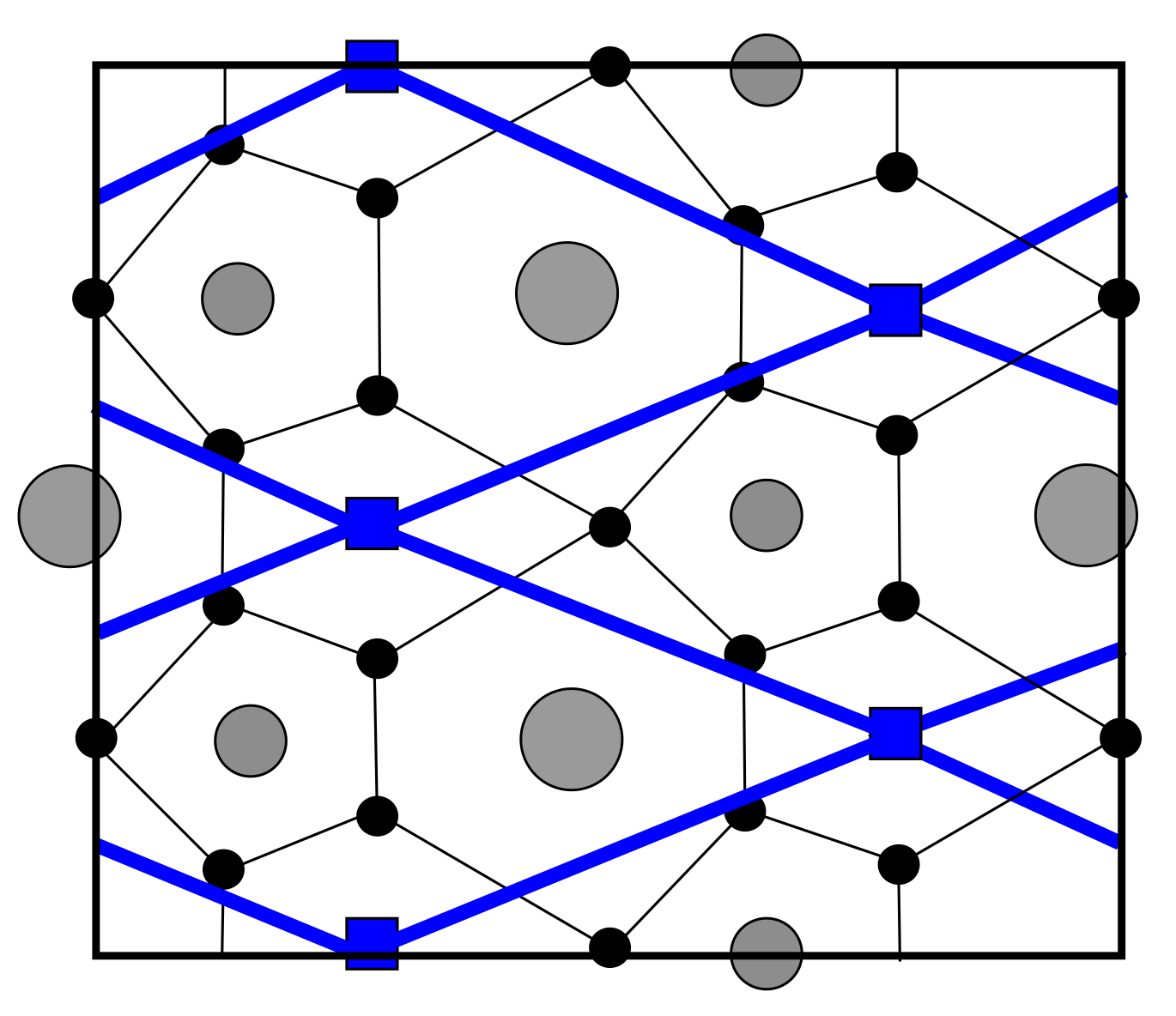}
	\caption{Diffusion lattice representation for the $\Sigma5$ grain boundary, in blue. Large gray circles denote Ni atoms on the GB plane, and small gray circles indicate Ni atoms on the next plan. Black dots mark all interstitial sites within the GB plane connected with the possible paths. Blue squares represent the effective sites used in the simplified diffusion graph, and blue lines indicate the idealized migration pathways between them.}
	\label{fig:S5_KMC_lattice}
\end{figure}

In the case of the $\Sigma5$ GB, the diffusion lattice graph is illustrated in Fig. ~\ref{fig:GB_graph}. This graph,  as the equivalent one for the $\Sigma3$ GB, has been constructed following the protocol illustrated in Fig.~\ref{fig:struct_to_graph}. However, in these cases, the possible migration jumps were first identified using a distance-based connectivity algorithm refined in the vicinity of the GB to reproduce the first-principles energetics of the corresponding structures reported in~\cite{DiStefano2015_Acta98_306}. For the $\Sigma3$ GB, minimal adjustments were required, whereas the more complex $\Sigma5$ boundary involved additional tuning. In the GB plane, excess volume leads to clusters of low-barrier sites which, in this work, are represented as a single effective site to simplify the graph while preserving dominant migration pathways and local connectivity. Fig.~\ref{fig:S5_KMC_lattice} illustrates the diffusion lattice for the $\Sigma5$ boundary. Then, spurious diffusion routes were eliminated based on geometric filtering, local-environment analysis, and manual validation.

The final GB structures were converted into multidirectional graphs after removing redundant Ni host sites. All graph construction and filtering were implemented using custom \texttt{pymatgen-based} scripts~\cite{Pymatgen_ONG2013314,Pymatgen_chemenv}.

The migration rates across sites, $i$ and $j$, were calculated from atomistic data using transition state theory (TST)~\cite{Mehrer2007}, as:

\begin{equation}\label{eq:arrhenius}
    k_{ij} = \nu_{ij} \exp\!\left(-\frac{\Delta E_{ij}}{k_\mathrm{B} T}\right),
\end{equation}
where $\nu_{ij}$ is the attempt frequency, $\Delta E_{ij}$ is the energy barrier, $k_\mathrm{B}$ is the Boltzmann constant, and $T$ is the absolute temperature. Time evolution was propagated using the rejection-free Bortz–Kalos–Lebowitz (BKL) algorithm~\cite{BORTZ197510}, ensuring statistically exact kinetics and high computational efficiency.

KMC trajectories were post-processed to extract macroscopic transport properties, following the Einstein-Smoluchowski equation \cite{Mehrer2007}. The mean-square displacement (MSD) was computed to evaluate direction-dependent effective diffusivities within the GB slab-model. These diffusivities serve as input parameters for the mesoscale diffusion model presented in the subsequent section.

\subsection{\label{sec:femD}Continuum modeling}

Finite element simulations to evaluate the effective hydrogen diffusivity in polycrystalline systems were performed using Ansys Mechanical~\cite{APDLRef}. The problem was formulated as a pure diffusion model governed by Fick’s second law. Unlike McNabb–Foster or Oriani-type trapping formulations~\cite{McNabbFoster1963, Oriani1970_trapping}, which explicitly introduce trap concentrations and kinetic exchange terms, the present framework captures analogous trapping and fast-transport effects implicitly through spatially resolved diffusivities derived from atomistic and KMC calculations. In this way, all microstructural heterogeneity is embedded directly in the diffusivity field rather than through additional constitutive parameters. Both model definitions, analyses, and post-processing have been automated using custom \texttt{pyansys}, an open-source Python interface for automating Ansys simulation workflows, scripts~\cite{pymapdl}. 

\subsubsection{Microstructure model}

\begin{figure}[!htbp]
    \centering
    \subfloat{%
        \includegraphics[width=0.49\columnwidth, frame]{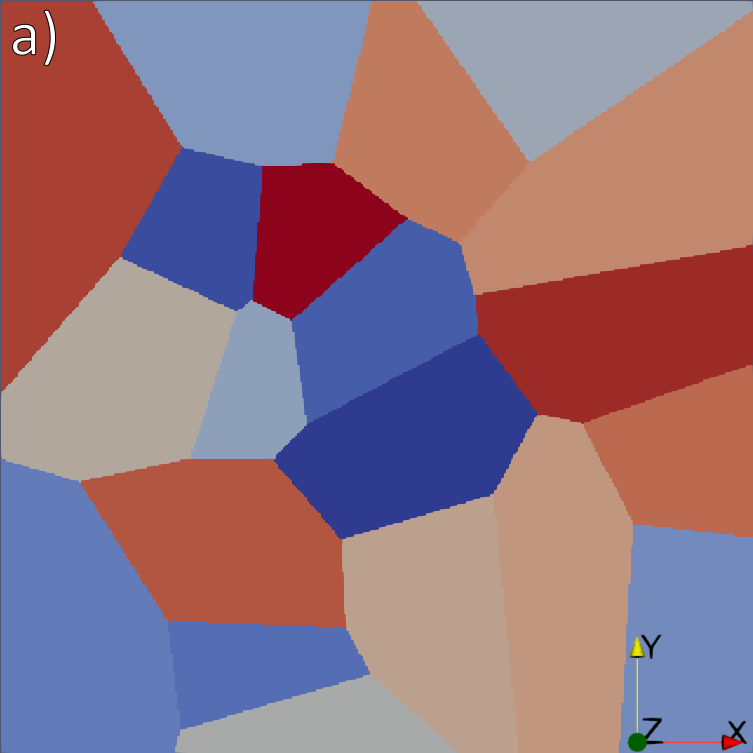}%
        \label{fig:microstructure_a}%
    }\hfill
    \subfloat{%
        \includegraphics[width=0.49\columnwidth, frame]{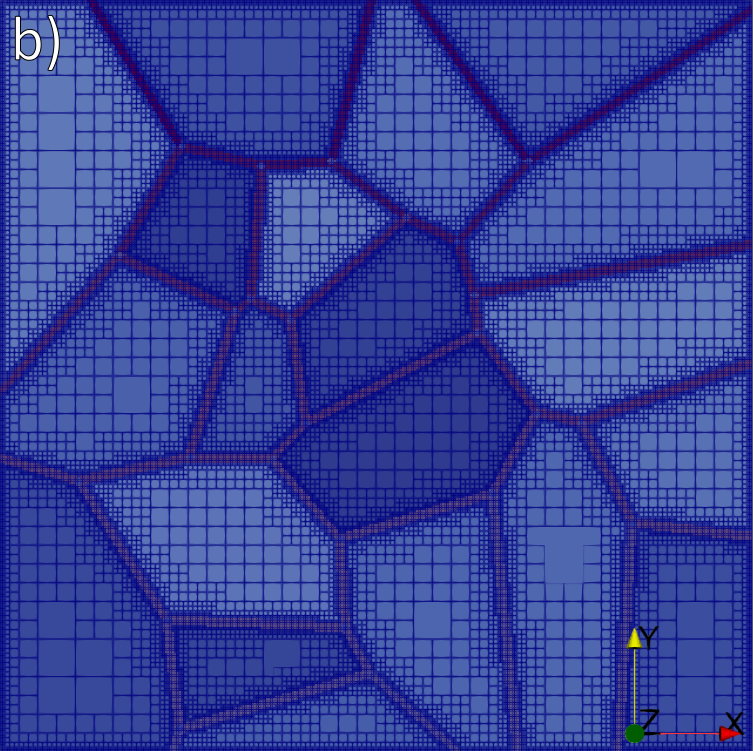}%
        \label{fig:microstructure_b}%
    }\\[0.5em] 
    \subfloat{%
        \includegraphics[width=0.49\columnwidth, frame]{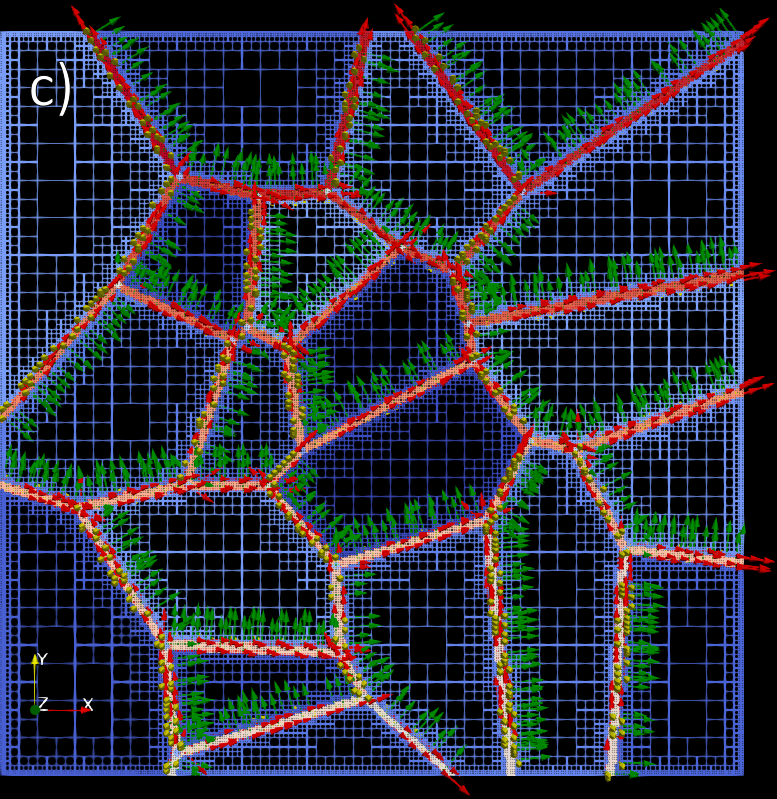}%
        \label{fig:microstructure_c}%
    }
    \caption{Continuum model preparation and setup: (a) generated microstructure, (b) microstructure with inflated GBs and TJs meshed using Octree meshing algorithm. Finally, (c) assigned local coordinate systems for anisotropic diffusivity directions.}
    \label{fig:microstructure_progression}
\end{figure}

Synthetic RVEs of polycrystalline materials were generated using an automated workflow. A voxel-based microstructure was first created using centroidal Voronoi tessellation, enabling control over grain count and average grain size (Fig.~\ref{fig:microstructure_a}). GBs and triple junctions (TJs) were identified from voxel connectivity and expanded into finite-thickness regions. TJs were modeled as an isotropic domain with enhanced diffusivity to reflect their experimentally observed fast-transport behavior~\cite{WANG200569, REDACHELLALI2013164}.

To reduce computational costs while preserving GB morphology, we used a 2.5D RVE, which is a slice with one voxel thickness, meshed with an adaptive Octree mesh~\cite{OctreeMesh, He_mesh}. The smallest element size was used in the GB and TJ regions, providing four elements across each boundary thickness, while grain interiors were coarsened; see Fig.~\ref{fig:microstructure_b}.

Grain interiors were assigned isotropic bulk diffusivity values~\cite{He_hydrogen}. GBs were treated as thin, anisotropic layers whose diffusivity tensors were mapped directly from KMC-derived transport properties. Local coordinate systems were created individually for each GB element by performing a principal component analysis on the voxel centroids defining the GB plane,~\ref{fig:microstructure_c}. This ensured proper alignment between the element orientation and the principal diffusivity directions (along and across the GB plane). Since we used a 2.5D representation for the diffusivities in the GB plane, we utilized the average of the two related diffusivities obtained from the KMC simulations. 
GB types ($\Sigma 5$ or $\Sigma 3$) were identified by the diffusivity tensors and were assigned randomly at the prescribed ratio. 

\subsubsection{Effective Diffusivity Analysis}
\begin{figure}[!htbp]
    \centering
    \includegraphics[width=0.69\columnwidth, frame]{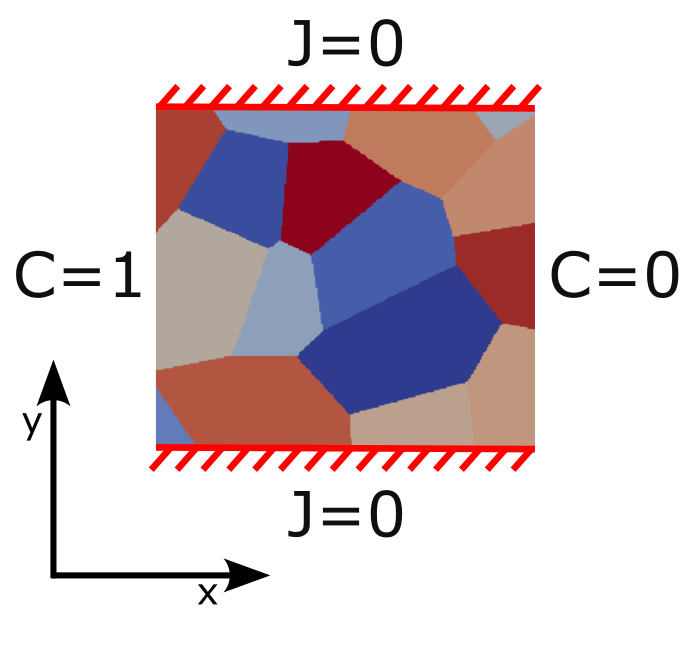}
    \caption{Boundary conditions for the permeation test applied to the 2.5D microstructure: normalize concentration $C=1$ on one face and $C=0$ on the opposite face, with zero flux ($J=0$) on the remaining boundaries. This setup enforces through-thickness transport, effectively reducing the problem to one-dimensional diffusion across the slab.}
    \label{fig:FEM_BC}
\end{figure}

Effective macroscopic diffusivity was determined by reproducing permeation tests. A constant hydrogen concentration was applied to one surface of the RVE, while the hydrogen flux through the opposing surface was monitored until a steady state was reached, see Fig.~\ref{fig:FEM_BC}. From the flux–concentration relationship obtained in this configuration, we can determine the effective diffusion coefficients for the polycrystalline system as a function of grain size and GB type distribution.

Although the microstructure model is 2.5D the boundary conditions enforce a concentration gradient only along the longitudinal axis, $x$ in Fig.~\ref{fig:FEM_BC}. This setup effectively reduces the transport problem to one-dimensional diffusion along that axis which can be described as: 

\begin{equation}
    J = -D_\mathrm{eff}^{1D} \frac{\Delta C}{\Delta x},
\end{equation}

where $J$ is the steady flux, $\Delta C$ is the imposed concentration difference, and $\Delta x$ is the RVE length. Assuming statistical isotropy of diffusion in a randomly oriented polycrystalline aggregate, the three-dimensional effective diffusivity can be obtained through:

\begin{equation}
    D_\mathrm{eff} = 3 D_\mathrm{eff}^{1D} = -3 J \frac{\Delta x}{\Delta C}.
\end{equation}

This approach connects the simulated steady-state permeation flux directly to the macroscopic hydrogen diffusivity of the modeled microstructure.

\section{\label{sec:results}Results}

\subsection{Verification of Kinetic Monte Carlo implementation}

Hydrogen diffusivity for bulk face-centered cubic Ni was first computed to benchmark the KMC implementation against analytical and experimental data. The diffusivity was found to be isotropic, consistent with cubic symmetry. As the simulation length increased, the diffusion coefficients along the main directions converged to identical values, confirming the accurate reproduction of isotropic diffusion. Details on convergence behavior are provided in the Supplementary Information. 

As expected, the KMC-derived diffusivities follow an Arrhenius relationship (Eqn.~\ref{eq:arrhenius}) with an activation energy $E_a = 0.37~\text{eV}$ and a pre-exponential factor $\nu_e = 5.48\times10^{12}~\text{s}^{-1}$~\cite{volkl_diffusion_1978}. Statistical uncertainty remained below 3\% across all temperatures, with convergence achieved by averaging approximately 500 trajectories of $10^6$ KMC steps each.

\begin{figure}[!htbp]
    \centering
    \includegraphics[width=0.99\columnwidth, frame]{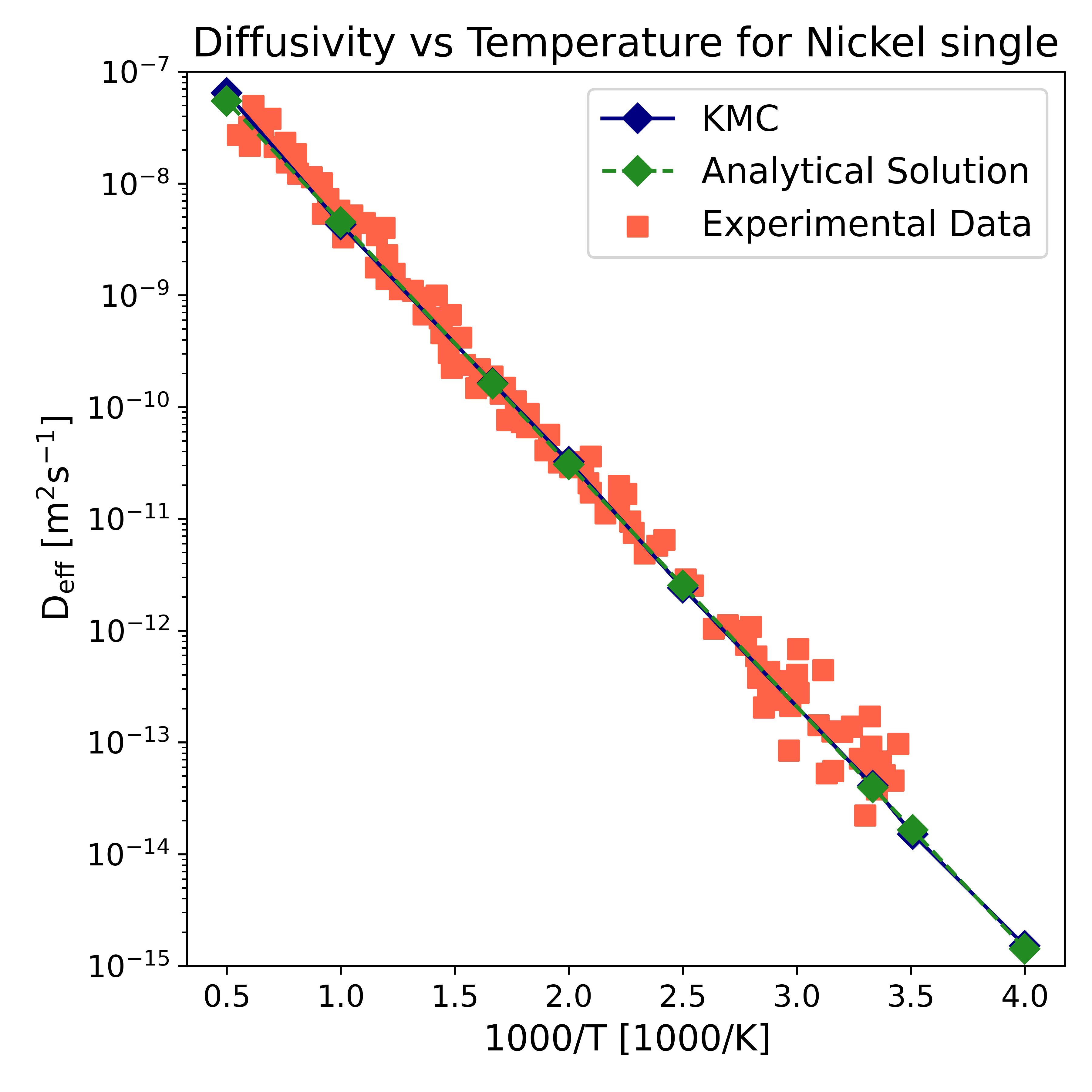}
    \caption{Temperature dependence of hydrogen diffusivity in bulk Ni. KMC results show excellent agreement with analytical and experimental data from Ref.~\cite{volkl_diffusion_1978}.}
    \label{fig:Ni_bulk_diffusivity}
\end{figure}

Figure~\ref{fig:Ni_bulk_diffusivity} compares the effective diffusivities from KMC, the analytical solution based on  TST, and experimental data~\cite{volkl_diffusion_1978}. The agreement is excellent, with deviations below 6.5\% from TST predictions. 

Having verified our implementation of KMC for bulk Ni and Fe, we next applied it to the GB slab-models.

\subsection{KMC simulation of Hydrogen Diffusion in proximity of Grain Boundaries}

\begin{figure*}[!htbp]
    \centering
    \subfloat{%
        \includegraphics[width=0.48\textwidth, frame]{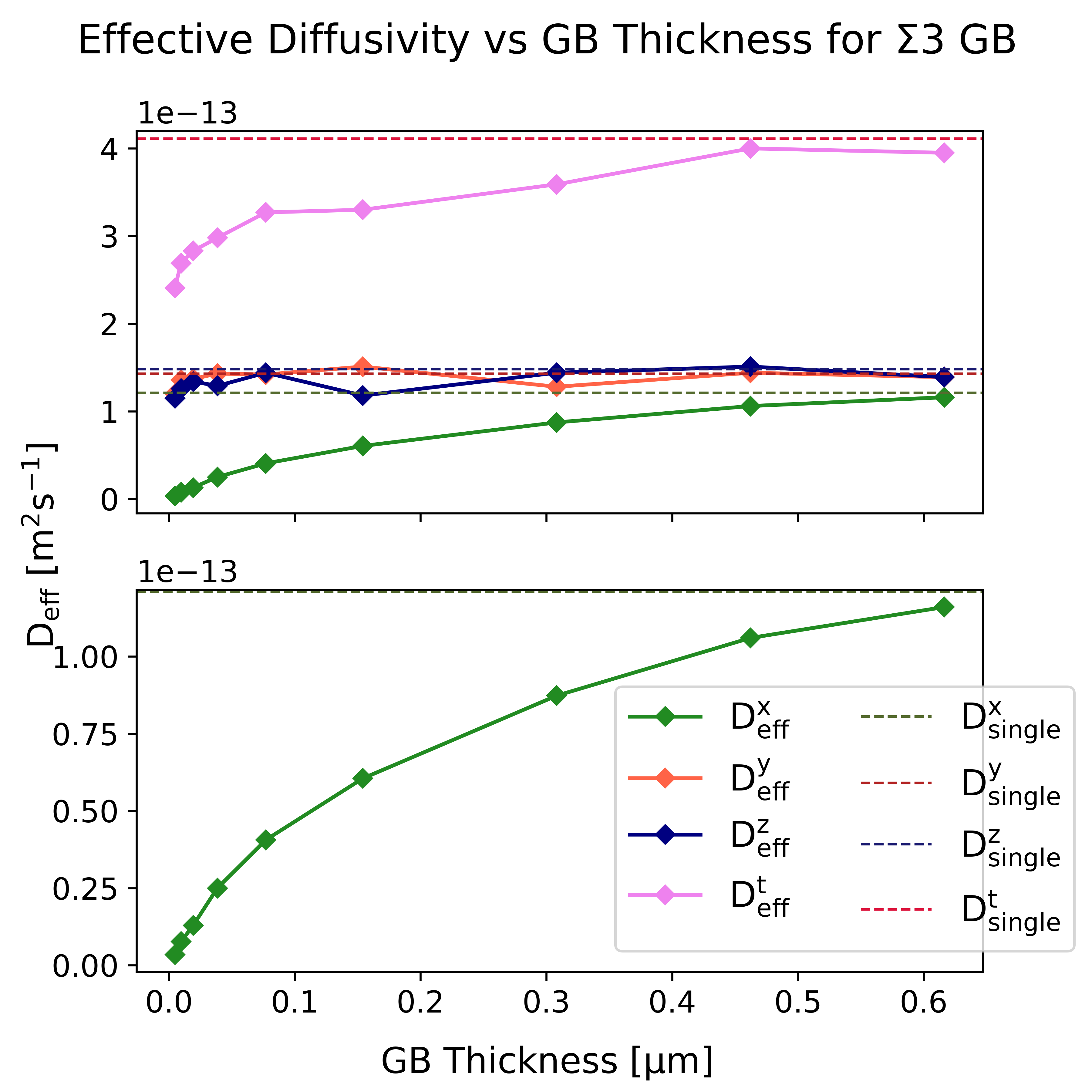}%
        \label{fig:Sigma3_diffusivity}%
    }\hfill
    \subfloat{%
        \includegraphics[width=0.48\textwidth, frame]{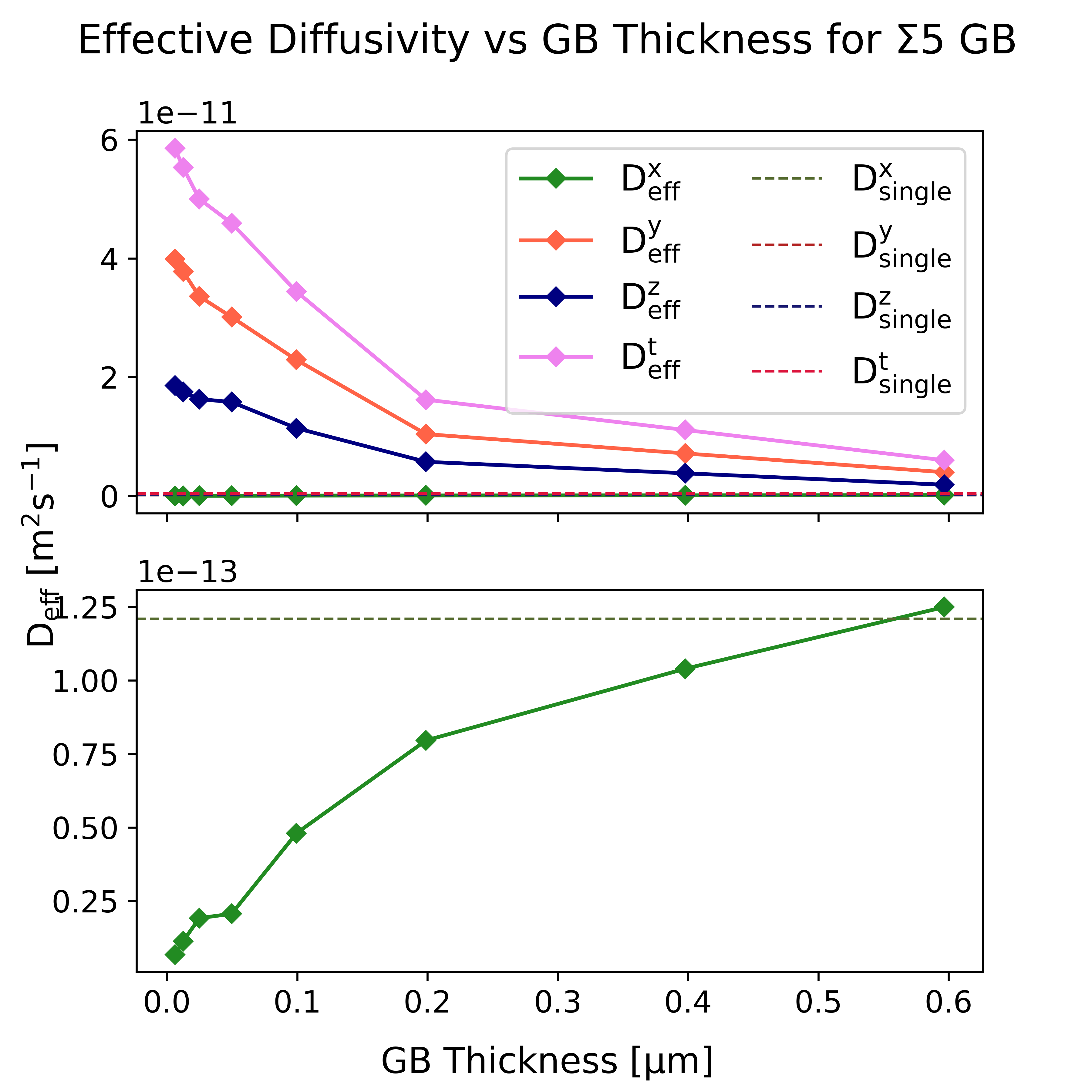}%
        \label{fig:Sigma5_diffusivity}%
    }
    \caption{Effective diffusion coefficients across ($D_x$) and along  ($D_y$ and $D_z$) the grain boundaries as functions of "GB thickness", i.e. distance between GB in the slab-model. The $\Sigma5$ boundary exhibits enhanced in-plane diffusivity and trapping-limited cross-plane transport, while $\Sigma3$ acts primarily as a two-dimensional diffusion barrier.}
    \label{fig:GB_diffusivity}
\end{figure*}

Effective diffusivities, $D_\text{eff}$, in slab geometries as a function of the distance between the repeated GBs were computed. Results are shown in Fig.~\ref{fig:GB_diffusivity}. Both $\Sigma5$ and $\Sigma3$ GBs exhibit orthotropic diffusivity. Consistent with atomistic simulation \cite{DiStefano2015_Acta98_306}, for the $\Sigma5$ GB the in-plane diffusivities ($D_x$, $D_y$) far exceed both cross-plane ($D_z$) and bulk values, reflecting low-barrier migration channels within the GB core. In contrast, the $\Sigma3$ GB shows slower diffusion in the cross-plane direction ($D_y$), consistent with its expected role as a structural diffusion barrier.

For both GB types, the diffusivities approach the bulk diffusivity $D_\text{bulk}$ as the domain size increases, consistent with the diminishing volumetric influence of GB regions. The $\Sigma 5$ GB exhibits some anisotropy in the in-plane diffusion direction due to longer jumps in the $y$ direction, despite similar energy barriers as the diffusion along the $z$ direction; see Fig. ~\ref{fig:S5_KMC_lattice}. 

\subsection{Finite Element Model Analysis}
The final step in the multiscale workflow involves finite element simulations of diffusion in polycrystalline models. These simulations were performed on eight microstructure models, spanning a range of average grain diameters, as summarized in Table~\ref{tab:RVE_sizes}. Examples of microstructure are reported in the Supplementary Information. 

\begin{table}[htp]
	\centering
	\caption{Representative Volume Element (RVE) dimensions used in the polycrystalline diffusion simulations for different average grain sizes.}
	\label{tab:RVE_sizes}
	\footnotesize  
	\setlength{\tabcolsep}{4pt}  
	\begin{tabular}{ccc}
		\toprule
		\textbf{Grain Size ($\mu$m)} & \textbf{RVE Size ($\mu$m)} & \textbf{GB Thickness ($\mu$m)} \\
		\midrule
		0.1   & 1 × 1     & 0.05 \\
		0.4   & 4 × 4     & 0.05 \\
		0.7   & 5 × 5     & 0.05 \\
		1.0   & 10 × 10   & 0.05 \\
		2.5   & 15 × 15   & 0.05 \\
		5     & 30 × 30   & 0.05 \\
		10    & 40 × 40   & 0.05 \\
		100   & 600 × 600 & 0.6  \\
		\bottomrule
	\end{tabular}
\end{table}

A grain boundary thickness of 0.025~$\mu m$ was used for all the RVEs, except for the 100~$\mu m$ grain RVE, where a GB thickness of 0.6~$\mu m$ was used. This choice was made to reduce computational costs, as GB thickness directly influences mesh resolution in our models. The number of grains ranged from 40 to 120 grains across all generated RVEs.

\begin{figure}[!htbp]
    \centering

    \subfloat{%
        \includegraphics[width=0.49\columnwidth]{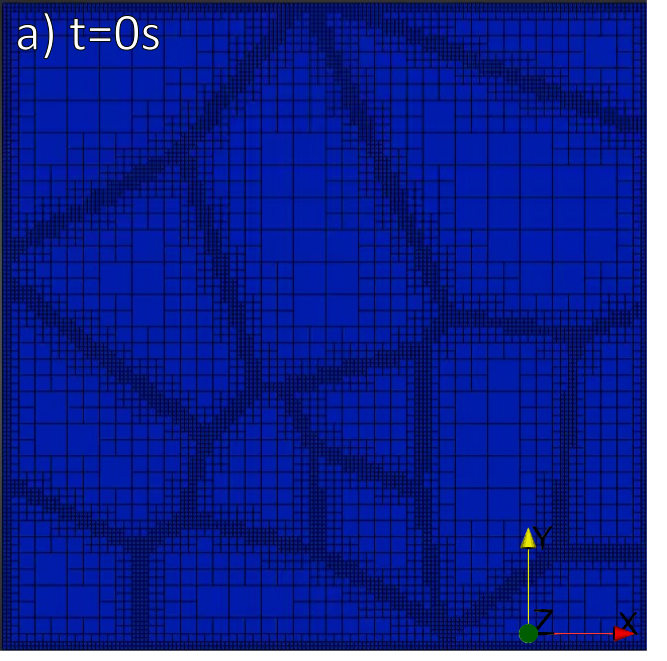}%
        \label{fig:conc_t0}%
    }\hfill
    \subfloat{%
        \includegraphics[width=0.49\columnwidth]{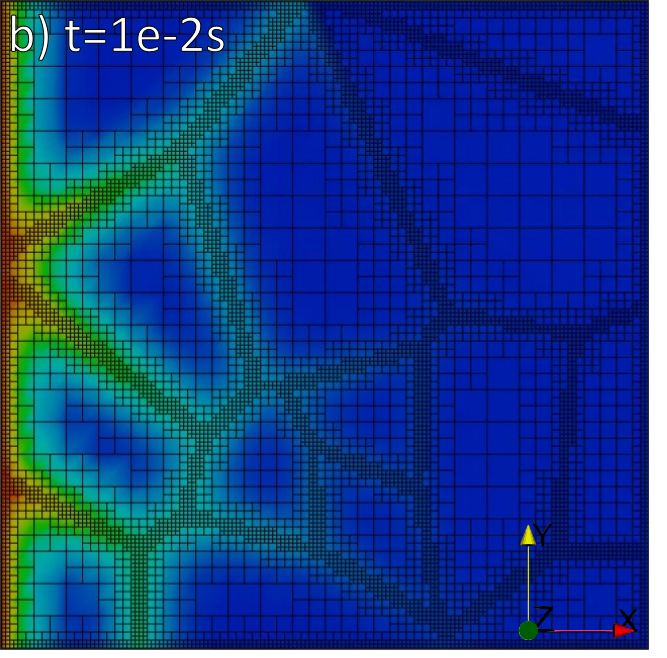}%
        \label{fig:conc_t1e2}%
    }\\[-2pt]

    \subfloat{%
        \includegraphics[width=0.49\columnwidth]{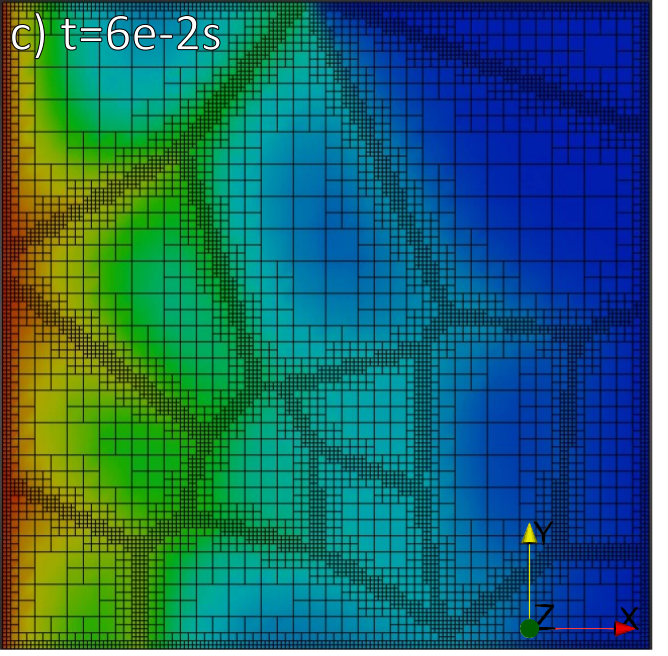}%
        \label{fig:conc_t6e2}%
    }\hfill
    \subfloat{%
        \includegraphics[width=0.49\columnwidth]{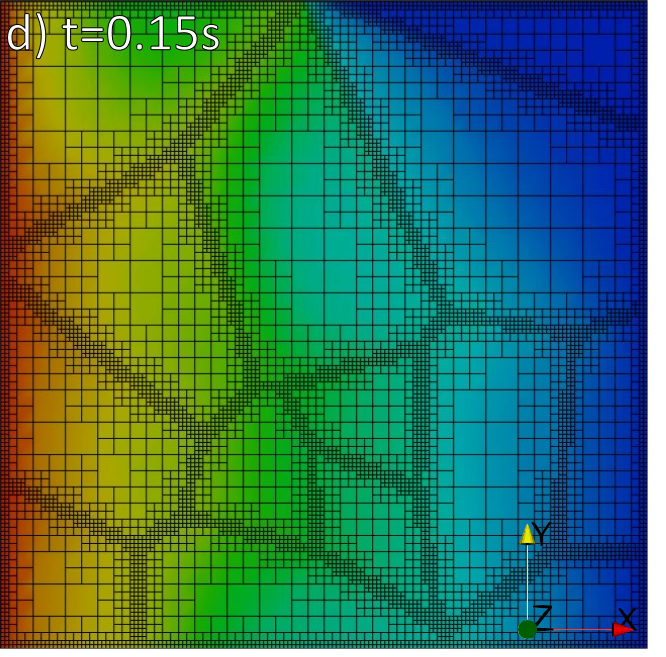}%
        \label{fig:conc_t015}%
    }

    \caption{Snapshots of the transient hydrogen concentration fields for an RVE with 1~$\mu m$ average grain size. The concentration fields indicate that the grain boundaries serve as dominant diffusion pathways.}
    \label{fig:FEM_conc_evolution}
\end{figure}

To illustrate the spatial evolution of hydrogen transport, Fig.~\ref{fig:FEM_conc_evolution} shows snapshots of the transient concentration fields in the 1~$\mu m$ grain RVE composed entirely of $\Sigma 5$-type grain boundaries. The simulations reveal rapid hydrogen migration along grain boundaries, which act as preferential pathways compared to the slower diffusion through grain interiors.

\begin{figure}[!htpb]
    \centering
    \includegraphics[width=0.99\columnwidth, frame]{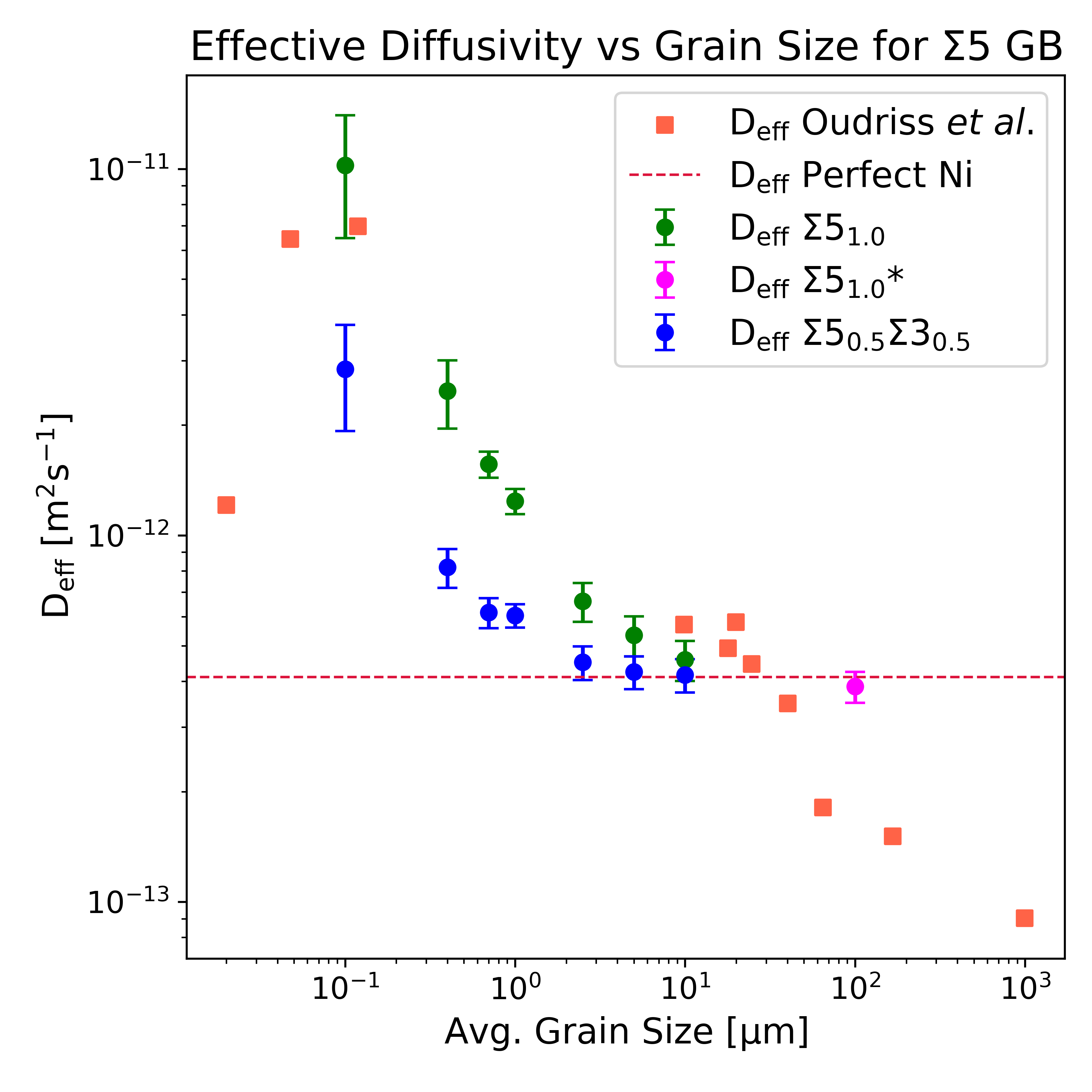}
    \caption{Effective hydrogen diffusivity ($D_\mathrm{eff}$) as a function of average grain size for polycrystalline Ni with $\Sigma5$ GBs. Symbols represent model predictions for different configurations: $\Sigma 5_{1.0}$ corresponds to microstructures where all grain boundaries are $\Sigma5$. $\Sigma 5_{1.0}^{*}$ denotes the 100~$\mu m$ grain size case where a larger GB thickness (0.6~$\mu m$) was used. $\Sigma 5_{0.5}\Sigma 3_{0.5}$ represents microstructures with 50\% $\Sigma5$ and 50\% $\Sigma3$ boundaries. Experimental data from Oudriss \textit{et al.}~\cite{OUDRISS20126814} are shown for comparison. The dashed line indicates the bulk Ni diffusivity for a perfect crystal. Refer to main text for detailed description and discussion}
    \label{fig:Deff_vs_grainsize}
\end{figure}

The effective diffusivity $D_\mathrm{eff}$, as a function of average grain size, is reported in Fig.~\ref{fig:Deff_vs_grainsize} as green circles. The $D_\mathrm{eff}$ for the structure with 100~$\mu$m grain size is shown in brown to indicate that it was obtained using a different GB thickness (see Table~\ref{tab:RVE_sizes}). These are microstructures consisting only of $\Sigma 5$-type GB. Experimental results for analogous random GB from~\cite{OUDRISS20126814} are reported as orange squares. All values were obtained at room temperature. 

Finally, microstructures with mixed grain boundary character, namely $50\% \text{ } \Sigma5$ and $50\% \text{ } \Sigma3$ GBs, are reported in Fig.~\ref{fig:Deff_vs_grainsize} as blue circles. These have intermediate effective diffusivities between the pure $\Sigma5$ case and the bulk limit.

Error bars reflect variability due to microstructural variations, different RVE creation instances, but do not account for other sources of uncertainty, such as parameter sensitivity or numerical discretizations.

\section{\label{sec:discussion}Discussion}

The results presented in the previous section demonstrate that hydrogen transport in polycrystalline nickel is strongly influenced by microstructural geometry and GB type. The multiscale framework developed in this work captures both trapping and fast-path effects through spatially resolved diffusivities derived from atomistic input, without relying on empirical trapping parameters or kinetic exchange terms.\\

While the qualitative behavior is consistent with atomistic studies, the KMC simulations enabled the direct extraction of homogenized anisotropic diffusivities suitable for continuum modeling. This step is essential for bridging atomistic insight with mesoscale transport predictions, particularly in microstructures where defect geometry and connectivity govern macroscopic behavior. 

Although the present workflow couples density-functional theory energetics with kinetic Monte Carlo (DFT+KMC) to obtain effective diffusivities in slab-geometry, an alternative route would involve molecular dynamics (MD) simulations, especially if leveraging machine-learning interatomic potentials (MLIPs) to achieve quantum-level fidelity ~\cite{ito2025,angeletti2025hydrogen}. Such approaches offer clear advantages in terms of automation and the ability to capture complex dynamical phenomena without manual intervention. However, the DFT+KMC strategy retains important benefits: it is significantly less computationally demanding and provides explicit control over individual migration events, enabling mechanistic interrogation of phenomena such as the origin of anisotropy within the GB plane or the contribution of trapping to effective diffusivity; see Supplementary Information for details. The main drawback of the DFT+KMC approach is its complexity and the effort needed to construct and validate both the atomistic input and the diffusion lattice, limiting full automation of the multiscale workflow. In contrast, MD combined with MLIPs offers a more streamlined workflow, but it comes at the expense of reduced interpretability and higher computational costs. These trade-offs highlight the complementary nature of the two strategies and suggest that hybrid approaches may offer an optimal balance between fidelity, efficiency, and mechanistic insight.\\

The predicted values of $D_\mathrm{eff}$ for microstructures composed exclusively of $\Sigma5$ grain boundaries (Fig.\ref{fig:Deff_vs_grainsize}), which we use as a model for random GB, exhibit good agreement with permeation measurements for fine- and intermediate-grained polycrystalline nickel (between 0.1 $\mu m$ and 10 $\mu m$ average grain sizes) reported by Oudriss et al.~\cite{OUDRISS20126814}, thus reinforcing the physical fidelity of the proposed multiscale framework. At the smallest grain size considered ($\sim$0.1~$\mu$m), this agreement should be interpreted cautiously, as this regime likely approaches the applicability limits of both the modeling assumptions and the experimental measurements, an aspect that was not examined in further detail in the present study. This level of correspondence is particularly noteworthy, given the deliberate simplifications introduced at each scale. These include reducing the diversity of grain boundary character to two representative prototypes ($\Sigma5$ and $\Sigma3$), employing a simplified Kinetic Monte Carlo lattice for the GB plane, representing grain boundaries as finite regions within the continuum model, and considering GB in-plane diffusivities to be isotropic. These approximations, while substantial, are intrinsic to multiscale modeling strategies, where systematic hierarchical simplification is essential for bridging several orders of magnitude in length-scale. The ability of the proposed approach to reproduce experimental trends under these constraints underscores its robustness and suitability for the quantitative analysis of microstructure-sensitive diffusion phenomena.

However, at the largest grain size considered (100~$\mu$m), the simulated $D_\mathrm{eff}$ approaches the bulk diffusivity of pure nickel while exceeding the corresponding experimental value. This deviation may be attributed to several factors. First, the continuum model assumes idealized GB connectivity and uniform diffusivity within GB regions, which may overestimate transport efficiency in coarse-grained microstructures where grain boundaries are sparsely distributed. Second, experimental measurements may reflect additional effects such as impurity interactions, sub-grain features, or defect populations. In this regime, hybrid approaches that combine microstructure-resolved GB transport with phenomenological trapping descriptions (e.g. McNabb–Foster or Oriani models) may therefore offer a viable modeling strategy. Third, at the constitutive level, while grain boundaries are represented as finite-thickness regions with tensorial diffusivity, a more accurate description would require the diffusivity to depend not only on orientation but also on the projection of the flux direction onto the local grain-boundary coordinate system. In other words, the effective diffusivity should differ depending on whether hydrogen is entering or leaving the GB region. This dependence on the flux direction would allow the model to more fully capture the concurrent trapping and fast-path transport behavior observed in atomistic and KMC simulations, particularly for boundaries such as $\Sigma5$ GB, but lies beyond the scope of the present study.

Microstructures containing mixed grain-boundary character exhibit intermediate transport behavior, confirming that macroscopic diffusivity is highly sensitive to the statistical distribution and connectivity of grain boundaries. Even partial replacement of fast boundaries with slower interfaces, represented in this work by the $\Sigma 3$, can markedly reduce global transport, illustrating that microstructural topology exerts a significant influence on hydrogen migration. This trend is consistent with observations in grain-boundary engineering, where increasing the fraction of special boundaries significantly improves resistance to hydrogen-assisted fracture~\cite{BECHTLE20094148}. Taken together, these results highlight that controlling hydrogen diffusion pathways is central to mitigating embrittlement. The present framework enables accurate prediction of diffusivity as a function of grain-boundary population, representing a critical step toward microstructure-informed alloy design strategies aimed at reducing hydrogen embrittlement.

\section{\label{sec:conclusions}Conclusions}
This work presents a multiscale modeling framework for hydrogen diffusion in polycrystalline nickel, integrating atomistic migration barriers, kinetic Monte Carlo simulations, and finite element modeling. By representing grain boundaries as finite-thickness regions with anisotropic diffusivities derived from atomistic input, the approach captures both trapping and fast-path effects without relying on empirical parameters. The model reproduces key trends observed experimentally, including the dependence of effective diffusivity on grain size and boundary character, and provides quantitative predictions that align well with permeation data for microstructures dominated by random grain boundaries.

The methodology enables the direct transfer of atomistic transport information into continuum-scale simulations, offering a physically grounded alternative to classical trapping-based models. 

While the present study focuses on nickel and specific GB configurations, the framework is general and can be extended to other materials and defect types, such as phase boundaries.\\

Future work will focus on incorporating temperature-dependent diffusivities and coupling the transport model with mechanical fields to study stress-assisted hydrogen migration. Also, a more refined treatment of directional trapping, where diffusivity depends not only on position and orientation but also on the flux direction, may further improve accuracy, particularly in coarse-grained microstructures. In addition, hybrid strategies that combine microstructure-resolved GB transport with classical phenomenological trapping formulations may offer a practical route to extend the model’s applicability across a wider range of grain sizes, an approach that has not been explored here. Finally, the integration of machine-learning interatomic potentials (MLIPs) represents a promising avenue for automating atomistic input generation and streamlining microstructure sampling. While this does not replace the interpretability offered by the current approach, it can significantly reduce manual effort and broaden the applicability of the framework, making it more accessible for both fundamental research and practical alloy design.

\section*{Acknowledgments}
The authors gratefully acknowledge \textbf{Prith Banerjee}, \textbf{Jay Pathak}, and \textbf{Ali Najafi} for their support throughout this work. We also thank \textbf{Pascal Salzbrenner} and \textbf{David Mercier} for their insightful discussions, which helped refine the methodology and interpretation of the results.

\subsection{Declaration of generative AI and AI-assisted technologies in the manuscript preparation process}

During the preparation of this work, the author(s) used Microsoft Copilot for grammar correction, syntax refinement, and improvements in clarity. After using this tool, the authors reviewed and edited the content as needed and take full responsibility for the content of the published article.

\clearpage
\bibliographystyle{apsrev4-2}
\bibliography{references}

\clearpage
\appendix
\onecolumngrid 
\setcounter{figure}{0}
\renewcommand{\thefigure}{S\arabic{figure}}
\setcounter{table}{0}
\renewcommand{\thetable}{S\arabic{table}}

\section*{Supplementary Information} 
\setcounter{figure}{0}
\renewcommand{\thefigure}{S\arabic{figure}}
\setcounter{table}{0}
\renewcommand{\thetable}{S\arabic{table}}

\subsection{\label{appendix:KMC_convergence}Convergence of KMC simulations}

\begin{figure}[htp]
    \centering
    \includegraphics[width=0.75\linewidth, frame]{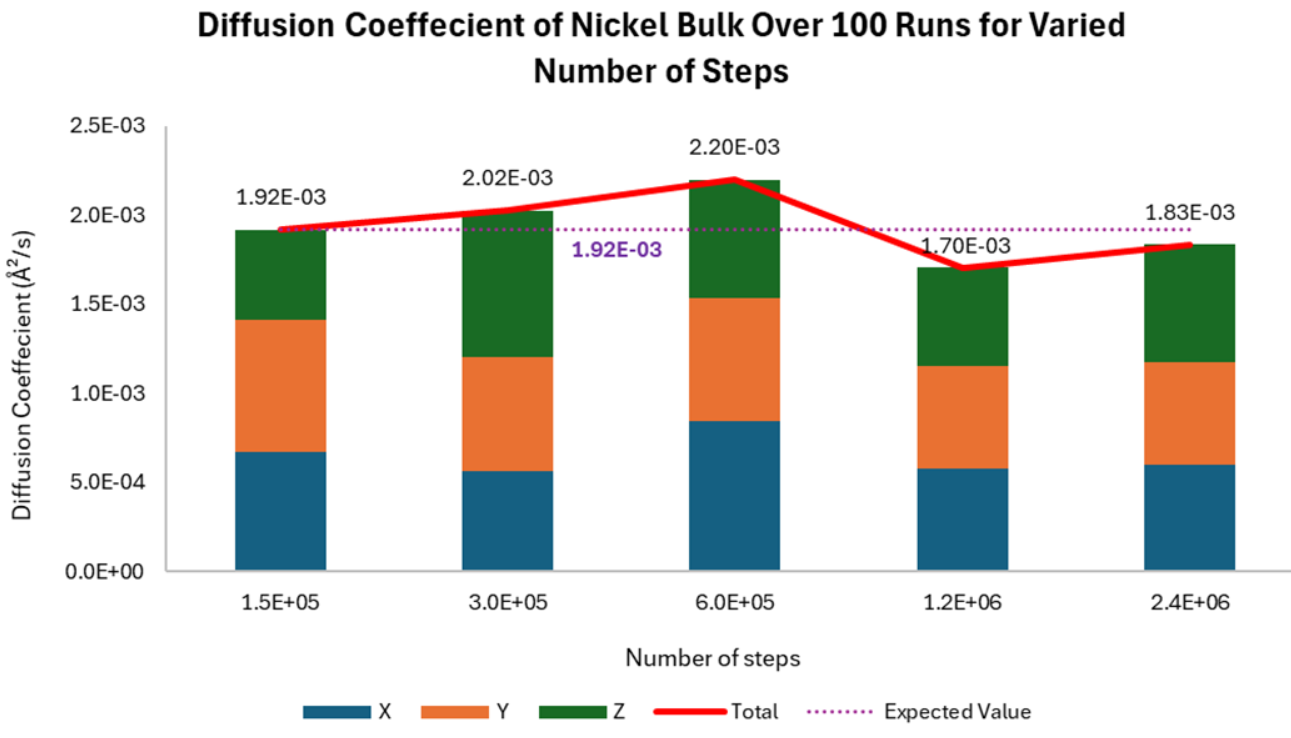}
    \caption{Convergence of diffusivity estimates with increasing simulation length. Increasing the number of KMC steps yields a more isotropic diffusion coefficient across the \(x\), \(y\), and \(z\) directions. Shown here are the averaged diffusion coefficients of bulk nickel over 100 independent simulations for varying numbers of steps.}
    \label{fig:Isotropic_Diffusion}
\end{figure}

\subsection{Comparison between KMC and simple analytical model}

To obtain effective diffusivities in slab geometries, we have used a Kinetic Monte Carlo (KMC) approach. A simple model [1] to calculate effective diffusivities in the presence of grain boundaries and grains is represented by the rule-of-mixture:

\begin{equation}
    D_\text{RoM} = f_\text{GB} D_\text{GB} + f_\text{Bulk} D_\text{Bulk},
\end{equation}
where $f_\text{GB}$ and $f_\text{Bulk}$ are the respective volume fractions.

\begin{figure}[htp]
    \centering
    \includegraphics[width=0.75\linewidth, frame]{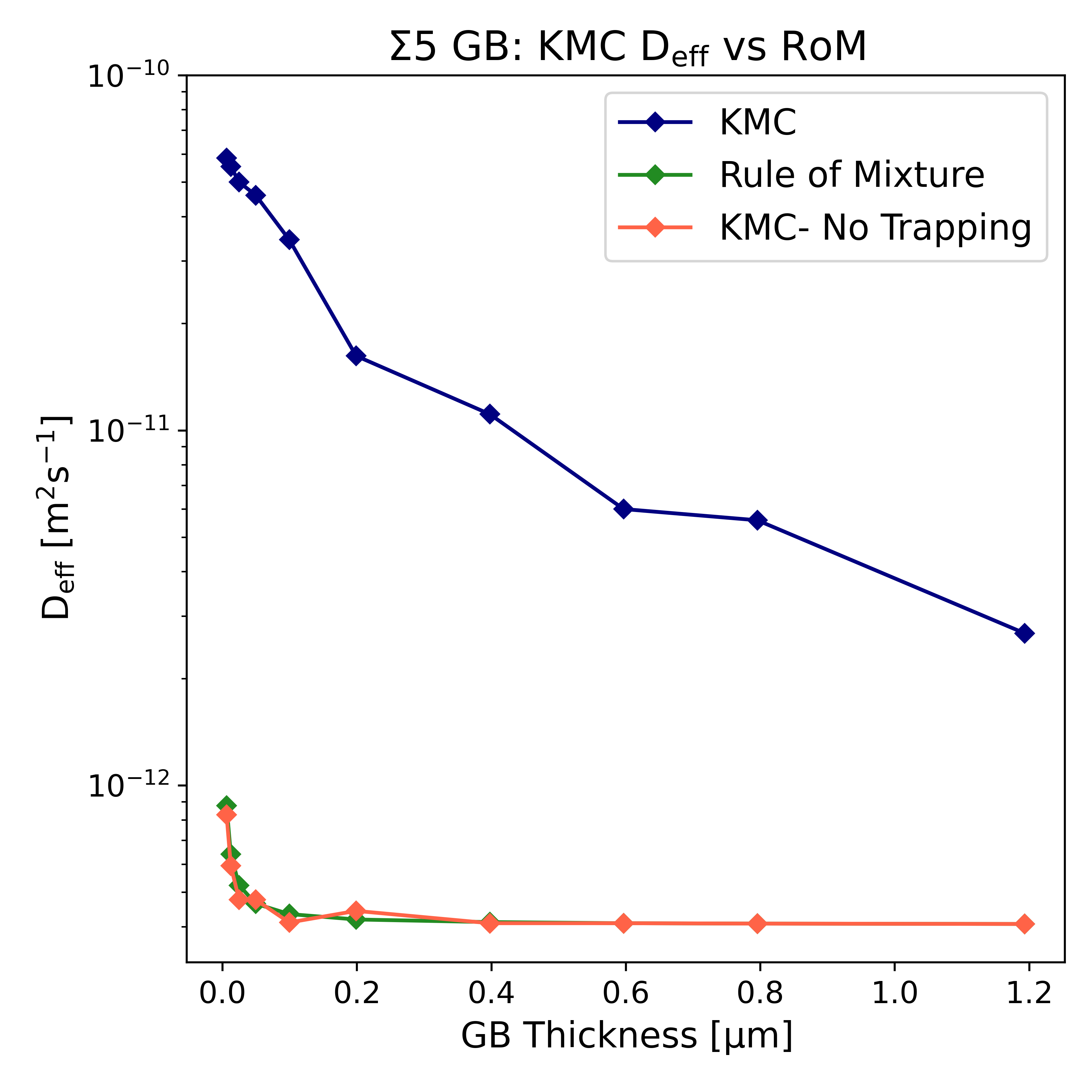}
    \caption{Comparison of KMC-derived $D_\mathrm{eff}$ with the rule-of-mixtures (RoM) estimate. The RoM underpredicts diffusivity due to neglected trapping and anisotropy; removing trapping effect restores agreement.}
    \label{fig:Sigma5_RoM_Diffusivity}
\end{figure}

The comparison of $D_\text{RoM}$ with effective diffusivity from KMC as a function of the GB distance in the slab-model is reported in Fig. ~\ref{fig:Sigma5_RoM_Diffusivity}. The rule-of-mixture results severely underestimated $D_\text{eff}$—by up to two orders of magnitude for $\Sigma5$—due to its neglect of interfacial anisotropy and trapping barriers. If we remove the trapping effects in the KMC model (by enforcing the probabilities to enter and leaving GB sites to be equal) the KMC and the rule-of-mixture results match. This confirms that the segregation effect is important and leads to faster diffusion. It is worth noticing that this was already understood and more refined models have been proposed [2].

\clearpage
\subsection{Examples of synthetic microstructures used in the main paper}

\begin{figure}[!htpb]
    \centering
    \includegraphics[width=0.75\linewidth, frame]{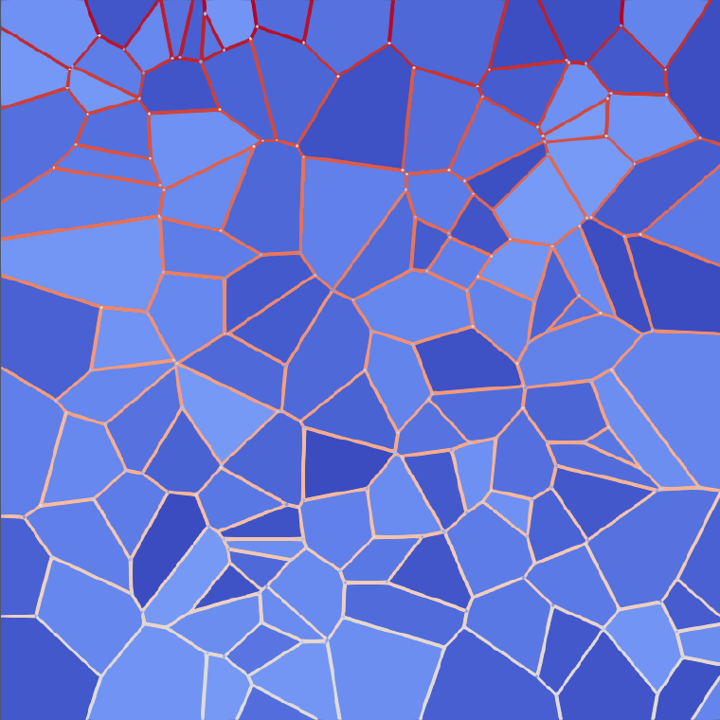}
    \caption{10×10 $\mu m$ RVE with 1$\mu m$ average grain size.}
    \label{fig:1um}
\end{figure}

\begin{figure}[!htpb]
    \centering
    \includegraphics[width=0.75\linewidth, frame]{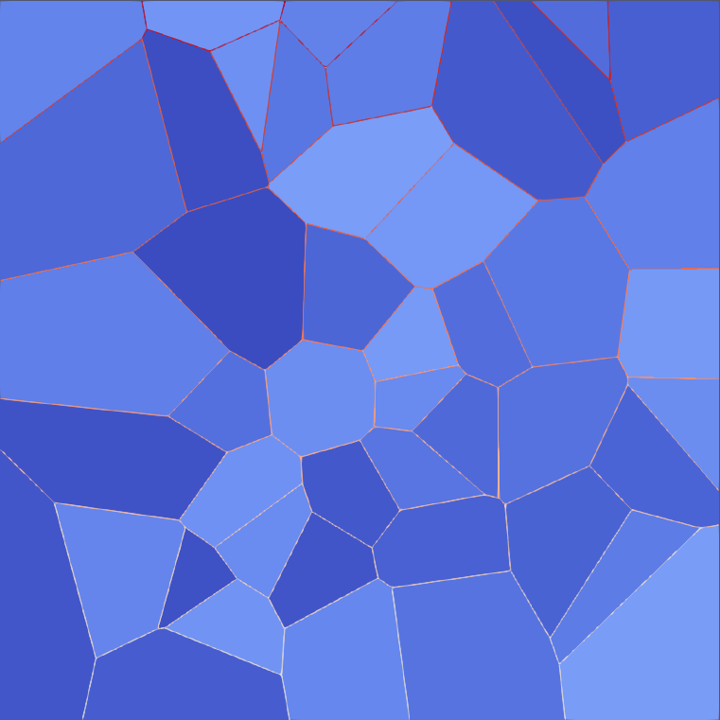}
    \caption{30×30 $\mu m$ RVE with 5$\mu m$ average grain size.}
    \label{fig:5um}
\end{figure}

\begin{figure}[!htpb]
    \centering
    \includegraphics[width=0.75\linewidth, frame]{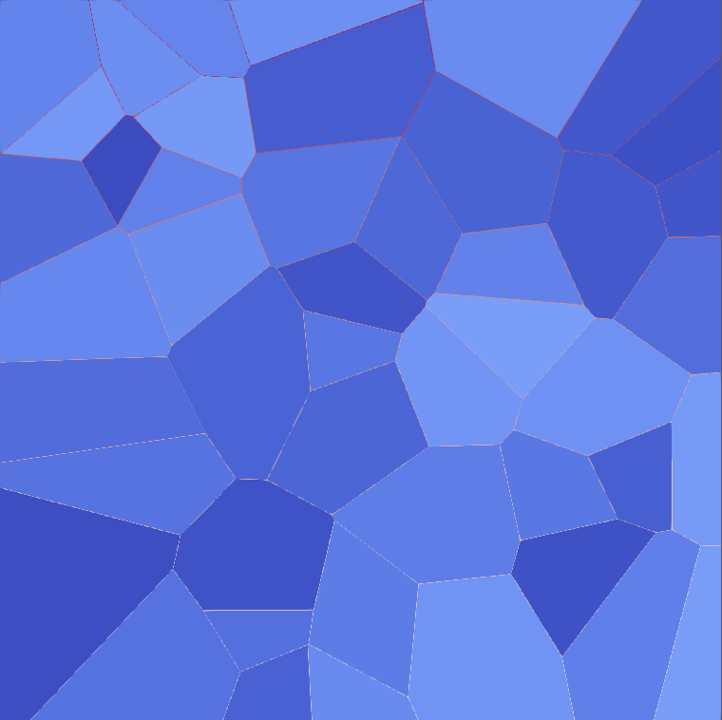}
    \caption{600×600 $\mu m$ RVE with 100$\mu m$ average grain size.}
    \label{fig:100um}
\end{figure}

\clearpage
\section*{\textbf{Supplementary References}}
\begin{enumerate}
	\item E. W. Hart, \textit{On the role of dislocations in bulk diffusion}, Acta Metallurgica, 5(10), 597 (1957).
	\item I.V. Belova and G.E. Murch, \textit{Diffusion in nanocrystalline materials}, Journal of Physics and Chemistry of Solids, 64(5), 873–878 (2003).
\end{enumerate}

\end{document}